\documentclass[preprintnumbers, floatfix, letterpaper, superscriptaddress,nofootinbib, twocolumn]{revtex4}
\usepackage[dvipdfmx]{graphicx}
\usepackage{amsmath}
\usepackage{amssymb}
\usepackage{subfigure}
\usepackage{hyperref}
\usepackage{url}
\usepackage{xcolor}
\usepackage{color}
\usepackage{mathrsfs}
\usepackage{amsfonts}
\usepackage{latexsym}
\usepackage{ragged2e}
\usepackage{epsfig}
\usepackage{textcomp}
\usepackage{phaistos}

\makeatletter
\renewcommand\@makefnmark{\hbox{\@textsuperscript{\normalfont\color{purple}\@thefnmark}}}
\renewcommand\@makefntext[1]{%
  \parindent 1em\noindent
            \hb@xt@1.8em{%
                \hss\@textsuperscript{\normalfont\@thefnmark}}#1}
\makeatother

\usepackage{caption}
\DeclareCaptionJustification{justified}{\leftskip=0pt \rightskip=0pt \parfillskip=0pt plus 1fil}
\def\beq{\begin{equation}}
\def\eeq{\end{equation}}

\def\mathbb{\Bbb}

\definecolor{vividviolet}{rgb}{0.62, 0.0, 1.0}
\definecolor{amaranth}{rgb}{0.9, 0.17, 0.31}
\definecolor{palatinateblue}{rgb}{0.15, 0.23, 0.89}
\definecolor{brightpink}{rgb}{1.0, 0.0, 0.5}
\definecolor{cornflowerblue}{rgb}{0.39, 0.58, 0.93}
\definecolor{deepcarminepink}{rgb}{0.94, 0.19, 0.22}
\definecolor{radicalred}{rgb}{1.0, 0.21, 0.37}
\colorlet{RED}{red}

\hypersetup{ linktoc=all,
    colorlinks, linkcolor={palatinateblue},
    citecolor={brightpink}, urlcolor={amaranth}
}

\graphicspath{{Images/}}

\renewcommand{\d}[1]{\ensuremath{\operatorname{d}\!{#1}}}

\graphicspath{{Images/}}
\renewcommand{\d}[1]{\ensuremath{\operatorname{d}\!{#1}}}
\makeatletter
\def\@fnsymbol#1{\ensuremath{\ifcase#1\or $\textleaf$ \or $\PHplaneTree$
\else\@ctrerr\fi}}%

\makeatother

\def\sideremark#1{\ifvmode\leavevmode\fi\vadjust{\vbox to0pt{\vss
 \hbox to 0pt{\hskip\hsize\hskip1em
 \vbox{\hsize1.5cm\tiny\raggedright\pretolerance10000
 \noindent #1\hfill}\hss}\vbox to8pt{\vfil}\vss}}}%
\begin{document}

\title{Holographic Consistency and the Sign of the Gauss-Bonnet Parameter}


\author{Yen Chin \surname{Ong}}
\email{ycong@yzu.edu.cn}
\affiliation{Center for Gravitation and Cosmology, College of Physical Science and Technology, Yangzhou University, \\180 Siwangting Road, Yangzhou City, Jiangsu Province  225002, China}
\affiliation{Shanghai Frontier Science Center for Gravitational Wave Detection, School of Aeronautics and Astronautics, Shanghai Jiao Tong University, Shanghai 200240, China}

\begin{abstract}

If Einstein-Gauss-Bonnet gravity is obtained as a low energy limit of string theory, then the Gauss-Bonnet parameter $\alpha$ is essentially the inverse string tension and thus necessarily positive. If one treats Einstein-Gauss-Bonnet gravity as a modified theory of gravity in the anti-de Sitter bulk in the bottom-up approach of holography, then there is no obvious restriction on the sign of the parameter \emph{a priori}, though various studies involving boundary causality have restricted the possible range of $\alpha$. In this short note, we argue that if holographic descriptions are to be consistent, then the Gauss-Bonnet parameter has to be positive. This follows from a geometric consistency condition in the Euclidean picture. From the Lorentzian signature perspective, black holes with a negative $\alpha$ lead to uncontrolled brane nucleation in the bulk and so the supposedly static geometry is untenable. In fact, even the ground state without a black hole is problematic. In other words, the bottom-up approach agrees with the top-down approach on the sign of the parameter. Some possible loopholes of the conclusion are discussed. 

\end{abstract}

\maketitle

\section{Introduction: Constraints on the Gauss-Bonnet Parameter}\label{1}

The Einstein-Gauss-Bonnet (EGB) gravity in $D$-dimensions in the presence of a negative cosmological constant $\Lambda$ is given by the action \cite{BD} (in the units $c=G=1$)
\begin{equation}
I_\text{EGB}=\frac{1}{16\pi} \int \text{d}^D x \sqrt{-g} \left(R-2\Lambda+\frac{\alpha}{D-4}\mathcal{G}\right),
\end{equation}
where $\mathcal{G}:=R_{\mu\nu\rho\sigma}R^{\mu\nu\rho\sigma}-4R_{\mu\nu}R^{\mu\nu}+R^2$ is the Gauss-Bonnet term, which is a topological invariant in $D=4$. Therefore we only consider EGB gravity in $D \geqslant 5$. 
The AdS curvature length scale $L$ is related to the cosmological constant via $\Lambda=-(D-1)(D-2)/L^2$.
We will refer to $\alpha$ as the Gauss-Bonnet parameter\footnote{Different authors have defined the Gauss-Bonnet parameter with different normalizations. For example, our $\alpha/(D-4)$ as a whole is very often refer to as the Gauss-Bonnet parameter/coupling constant \cite{BD}. Evidently it has dimension length squared. In another convention, the coupling constant is defined as $\lambda_\text{GB}L^2/[(D-3)(D-4)]$, see, e.g., \cite{0911.3160}. The $(D-4)$ factor in the denominator allows Glavan and Lin to construct the so-called ``novel'' $D=4$ EGB theory \cite{1905.03601}, whose status is controversial due to a number of subtleties and issues \cite{2003.11552,2004.02858,2004.03390,2004.12998,2005.03859,2005.12292,2009.10715}. We will not specifically discuss this theory in this work, though our analysis would apply to this case just as well.}.
This theory can be interpreted as a higher curvature correction to general relativity (GR), and in fact can be derived as a low energy limit of heterotic string theory \cite{BD,Z,GW,GS}, in which the Gauss-Bonnet parameter is proportional to the inverse string tension, or related to the coupling constant $\alpha'$ in the string worldsheet action \cite{Z}. From this ``top-down'' view, $\alpha$ is clearly positive. Black holes in the EGB theory have been studied in details, see for example, \cite{cai, 0504127, 0504141}.

In gauge/gravity correspondence (hereinafter, ``holography''), it has become a common practice to employ modified gravity theories in the anti-de Sitter (AdS) bulk to model field theories with certain desired properties. Strictly speaking, the bulk physics is described by string theory. However, under certain well-defined circumstances (the string coupling, and the string length to AdS length scale ratio, are small), one hopes that stringy effects can be neglected. If so, then it suffices to deal with (mostly) semi-classical gravity. In such a ``bottom-up'' approach, one can often find analytic black hole solutions to work with. The hope is that such effective theories can in fact be embedded into string theory. For examples, attempts have been made to embed massive gravity \cite{1807.00591}, bimetric gravity \cite{2106.04614}, Ho\v{r}ava gravity \cite{1211.0010} and Lifshitz gravity \cite{1512.03554} in string theory, which provide some justifications for their applications in the literature. However, not all effective field theories can be embedded into string theory, as expressed by the ``Swampland Conjecture'' \cite{0605264,1903.06239}.

On the other hand, from the bottom-up perspective, the Gauss-Bonnet parameter is just a coupling constant. It is not obvious that it \emph{has} to be positive. 
This is not to say that there is no constraint at all for the range of $\alpha$. For example, assuming that $\alpha>0$, various causality considerations (including: scattering amplitude, requirement that graviton Shapiro time delay as opposed to time \emph{advance}, shock wave characteristic analysis\footnote{In EGB theory, there exist superluminal propagation of gravitons along
the characteristic surfaces of the PDEs, which determines the causal cone of the theory, distinct from the null cone \cite{1406.0677,1406.3379}.}, hyperbolicity, quantum entanglement constraints) \cite{1407.5597,1508.05303, 2101.02461,1401.5089} require $\alpha$ to be small. By small, we mean $\alpha \ll \ell_P^2$, the square of the Planck length. This is not automatically satisfied in string theory: in the weakly coupled case we can have $\alpha \gg \ell_P^2$. However, the existence of higher spin massive particles in the full string theory protects causality \cite{1407.5597}. From these constraints, $\alpha<0$ is not ruled out, as shown by the hyperbolicity and boundary causality analysis in \cite{1610.06078}. (See also \cite{0802.3318,0906.2922,0911.4257,0911.3160} for related discussions and constraints.) 
Boundary causality, as stated in the Gao-Wald theorem \cite{0007021}, essentially forbids signals to take a short-cut through the AdS bulk, so in particular, points which are not causally related on the boundary cannot be causally related through the bulk\footnote{In the ``novel'' $D=4$ case, Ge and Sin \cite{2004.12191} argued on causality ground that $\alpha < 0$.  In the actual Universe, observations prefer $\alpha>0$ \cite{2006.15017}.} (see also \cite{1604.03944} and a recent lecture note by Witten \cite{1901.03928}). The Gauss-Bonnet parameter $\alpha$ has effects on the thermodynamics of black holes and thus on the dual field theory. In particular the Kovtun-Son-Starinets (KSS) viscosity bound \cite{0104066,0309213}
is modified \cite{0712.0805,0712.0743,0812.2521}. Interestingly, such an entropy bound is closely related to the aforementioned causality issue \cite{0808.1919, 0904.4805}. Furthermore, eikonal instability develops if $\alpha$ is not small enough \cite{1701.01652,1705.07732}.

Various applications of EGB black holes in holography had been studied, see, e.g., \cite{1103.3982,1305.4841, 1343004, 1311.3053, 1508.03364,1608.04208,1712.02772,Lu,2101.00882,0407083,1610.08987,2106.13942,2102.12171,1512.05666}. In fact the Gauss-Bonnet term is closely related to the central charges of the conformal field theory \cite{0906.2922,0712.0743}.
Pathological behavior in the evolution of entanglement entropy and mutual information in the case of $\alpha<0$ (the mutual information becomes discontinuous during thermalization) did led Sircar et al. \cite{1602.07307} to conjecture that EGB theory with $\alpha<0$ may be inconsistent\footnote{\label{ft5}Another reason to disfavor $\alpha <0$ is that under Hawking evaporation the end state would be a naked (null) singularity with a finite mass, thus violating weak cosmic censorship. One can of course argue that at such a scale new higher order terms in the action would be important and the physical picture might change. Nevertheless, $\alpha>0$ case has no such problem -- the black hole tends to a zero mass and zero temperature state under Hawking evaporation \cite{2103.00257}, at least in 5-dimensions.}. In view of these, we raise the following question: \emph{is EGB gravity with $\alpha<0$ a consistent effective theory, with some possible embedding into string theory, or does it belong to the Swampland?} We would argue in favor of the latter case.

Note that in the asymptotically flat case, working in the effective theory regime, Cheung and Remmen have shown that $\alpha$ is nonnegative for any unitary tree-level ultraviolet completion of the Gauss-Bonnet term, which is free of ghosts or tachyons \cite{1608.02942}. Our work, which utilizes a completely distinct method, provides strong evidence for $\alpha \geqslant 0$ in the AdS case as well. 

\section{Holographic Consistency and Brane Nucleation}\label{II}

Holography relates gravitational theories in the AdS bulk to non-gravitational field theory on the boundary (see \cite{1501.00007,1409.3575} for a review). Such a surprising relation between two radically different types of theory is highly nontrivial and requires consistency conditions to be valid. A particularly deep condition of this kind in the Euclidean formulation that relates the minimum of a probe brane action to a gravitational bulk action \cite{1310.6788,1311.4520,1411.1887,1602.07177}:
\begin{equation} 
S_g^*=\frac{N}{\gamma}S_b^*,
\end{equation}
where $S_g^*$ is the on-shell gravitational action in the bulk, $N$ and $\gamma$ are the number of colors and the scaling exponent for the free energy of the boundary field theory, respectively, and $S_b^*$ is the probe brane action. 
Surprisingly, this consistency condition holds if a certain ``isoperimetric \emph{inequality}'' \cite{1411.1887,1610.07313} (the superscript ``E'' denotes a Euclidean quantity; likewise superscript ``L'' denotes a Lorentzian one -- not to be confused with the AdS curvature length scale $L$) 
\begin{equation}\label{brane}
\mathfrak{S}^\text{E}:=A(\Sigma)-\frac{D-1}{L}V(M_\Sigma) \geqslant 0,
\end{equation}
holds for any co-dimension 1 hypersurface $\Sigma$ in the bulk homologous to the boundary. Here $A(\Sigma)$ and $V(M_\Sigma)$ denote the area and the volume enclosed by $\Sigma$, respectively.
We see that $\mathfrak{S}^\text{E}$ measures the competition between the area and the volume.
Up to a brane tension constant coefficient, $\mathfrak{S}^\text{E}$ is the action of a probe BPS (Bogomol'nyi–Prasad–Sommerfield) brane that wraps around the black hole at a constant coordinate radius $r$.
If the Wick-rotated spacetime is an Einstein manifold, this inequality is related to the topology at infinity and the Yamabe invariant, as discussed in the theorem of Wang \cite{Wang1,Wang2}. In holographic settings, many manifolds are not Einstein, and so inequality (\ref{brane}) must be checked on a case by case basis. As mentioned in \cite{1504.07344}, generalizations to Wang's theorem do exist -- for example, the works of Witten-Yau \cite{WY} and Cai-Galloway \cite{CG} -- but they are in general not useful in deciding whether inequality (\ref{brane}) holds. The reason inequality (\ref{brane}) is referred to as an ``isoperimetric inequality'' can be appreciated as follows \cite{1610.07313}: in planar geometry, the area bounded by a closed curve satisfies $A\leqslant {\ell^2}/{4\pi},$
where $\ell$ is the length of the curve, and with equality holds if and only if the closed curve is a circle. This is the usual isoperimetric inequality. We can call this a ``consistency condition'' for a geometry to be embedded in $\textbf{R}^2$, i.e. a closed curve cannot bound too large a volume, otherwise it ceases to be planar. Likewise, inequality \ref{brane} is a geometric consistency condition for asymptotically hyperbolic manifolds, i.e., the Euclidean signature version of AdS spacetime employed in holography (we refer the readers to \cite{1610.07313} and \cite{1504.07344} for the details and subtleties). 

The Lorentzian version of inequality (\ref{brane}) has a clear physical picture: if $\mathfrak{S}^\text{L}$ becomes negative at large $r$, the spacetime will nucleate a copious amount of branes via a Schwinger-like process in the bulk (see, for example, \cite{1005.4439} and the Appendix of \cite{2005.12075}). This would back-react on the metric, which in turn means that the original spacetime cannot be static. Any static black hole solution in classical gravity theories that suffer from this ``Seiberg-Witten instability'' \cite{9903224,0409242} is therefore \emph{not} a consistent object in the bulk\footnote{This kind of brane nucleation instability was also employed to investigate ``Euclidean wormholes'' (solutions with multiple boundaries) in the classic work of Maldacena-Maoz \cite{0401024}, and the recent follow-up by Marolf-Santos \cite{2101.08875}. In the context of \cite{0912.1061} such instability was dubbed ``Fermi Seasickness''. See also \cite{1910.06348,2104.00022}.}. Note that as with the Schwinger pair-production, this effect is non-perturbative \cite{0409242}. We remark that it is possible for $\mathfrak{S}^\text{E}$ to be everywhere non-negative but $\mathfrak{S}^\text{L}<0$ at some locations. Holography should be regarded as fully consistent when both inequalities are valid (although $\mathfrak{S}^\text{L}<0$ is acceptable if the boundary system only exists for a sufficiently short period so that any Lorentzian pathology would not have the time to influence the black hole geometry). See \cite{1504.07344} for detailed discussions.

A classic example is that of toral Reissner-Nordstr\"om black holes \cite{0905.1180,0910.4456,1012.4056}. Although they are valid solutions to Einstein-Maxwell gravity for all values of the charge $Q$, they are not a consistent solution in the bulk when the electrical charge becomes near-extremal (for $\text{AdS}_5$, this occurs at $0.958 \times Q_{\text{extremal}}$ \cite{1012.4056}). This makes sense from the dual field theory (quark-gluon plasma) perspective, since no plasma can be arbitrarily cold. Another example is provided in \cite{1008.0231}, in which the brane action is used to explain why ``squashed'' AdS$_5$-Schwarzschild black hole is not consistent in the aforementioned sense, thus explaining why they do not have a sensible field theory dual, a phenomenon which was observed in \cite{0901.2574}. Interestingly, the same method predicts a bound on the strength of magnetic field compared to temperature squared of the surrounding plasma (independent on the AdS length scale $L$): $B \leqslant 2\pi^{3/2}T^2$; and cosmic magnetic fields during the plasma epoch of the early Universe may well come close to reaching that bound \cite{1409.3663}.

Note that one should not be too alarmed by the word ``inconsistency'' used in this context. Often what happens -- as illustrated in the examples above -- is that there is no sensible gravitational dual of a field theory precisely because the field theory description is also problematic (e.g. no such thing as an arbitrarily cold plasma). Thus the gauge/gravity duality holds and is consistent in the sense that both sides of the descriptions fail. What is meant by ``inconsistent'' in the sense of \cite{1411.1887,1504.07344} is that the relevant geometry would not satisfy the isoperimetric inequality. Thus one must be careful to distinguish between inconsistency in this very precise sense, from the broad consistency of the gauge/gravity correspondence as a matter of principle. In this work, holographic consistency always refers to the former. 

\section{Brane Nucleation of Einstein-Gauss-Bonnet Black Holes}
We will show that asymptotically locally AdS static black holes in EGB theory is not consistent in the aforementioned sense, even for the neutral case (for which $\mathfrak{S}^\text{E}$ is equivalent to $\mathfrak{S}^\text{L}$). 

The black hole solution of EGB gravity in $D$-dimensions is \cite{cai, 0504127, 2004.12191}
\begin{equation}
\d s^2 = -f(r) \d t^2 + f(r)^{-1} \d r^2 + r^2 \d \Omega_{k,D-2}^2,
\end{equation}
where $\d \Omega_{k,D-2}^2$ is a metric on a $(D-2)$-dimensional manifold $\Sigma_{k,D-2}$ with constant sectional curvature $k=\{-1,0,+1\}$, and 
\begin{equation}\label{metricfn}
f(r)=k-\frac{r^2}{2\alpha (D-3)}\left[-1+\sqrt{1-4\alpha(D-3)g(r)}\right],
\end{equation}
with 
\begin{equation}
g(r)=\frac{1}{L^2}-\frac{16\pi M}{(D-2)\Omega_{D-2}r^{D-1}},
\end{equation}
where $M$ is proportional to the black hole mass.
Here $\Omega_{D-2}$ denotes the unit area of $\Sigma_{k,D-2}$. For example, for a 3-sphere, $\Sigma_{1,3}=2\pi^2$, whereas for a cubic torus with periodicity $2\pi K$ on each copy of its $S^1$, we have $\Sigma_{0,3}=8\pi^3K^3$. In the presence of a Maxwell field, $g(r)$ would contain an additional charge term $2Q^2/[(D-3)(D-2)r^{2(D-2)}]$; its inclusion does not change the qualitative results discussed below.

The brane action $\mathfrak{S}^E$ is evaluated as follows:
\begin{flalign}
\mathfrak{S}^\text{E}&:=r^{D-2} \int \d \Omega_{k,D-2} \int \d \tau \sqrt{g_{\tau\tau}}\\ \notag &- \frac{D-1}{L} \int \d\tau \int \d \Omega_{k,D-2} \int^r_{r_h} \d r' r'^{D-2} \sqrt{g_{\tau\tau}g_{r'r'}},
\end{flalign}
where $g_{\tau\tau}$ is the Wick-rotated metric coefficient $g_{tt}$. For the uncharged case $g_{\tau\tau}=f(r)$. 
Here $r_h$ denotes the (Euclidean) event horizon.
(In fact, the area and volume terms correspond to the Dirac-Born-Infeld (DBI) term and Chern-Simons term, this can potentially generalize the consistency conditions to a more general backgrounds \cite{1411.1887}).
The normalized time coordinate $t/L$ is periodically identified with a periodicity $2\pi P$, so 
\begin{equation}
\mathfrak{S}^\text{E}= 2\pi P L \Omega_{D-2} \left[\underbrace{r^{D-2}\sqrt{g_{\tau\tau}}-\frac{r^{D-1}-r_h^{D-1}}{L}}_{=:\mathfrak{F}(r)}\right].
\end{equation}

The sign of $\mathfrak{S}^\text{E}$ depends on the competition between the terms in $\mathfrak{F}(r)$. For a sufficiently large $r$ (so that the constant term involving $r_h$, and the sectional curvature $k$ can be neglected), it is straightforward to show that for $\alpha < 0$, the bracket terms grow like 
\begin{flalign}
\mathfrak{F}(r)\sim&\frac{r^{D-1}}{\sqrt{2|\alpha| (D-3)}}\left[\left(\sqrt{1+\frac{4|\alpha|(D-3)}{L^2}}-1\right)^{\frac{1}{2}}\right.\\ \notag
&\left.-\frac{\sqrt{2|\alpha|(D-3)}}{L}\right], 
\end{flalign}
which is negative by elementary algebra (similarly, the corresponding quantity for $\alpha>0$ case is positive). 

\begin{figure}[h!]
	\centering
	\includegraphics[scale=0.40]{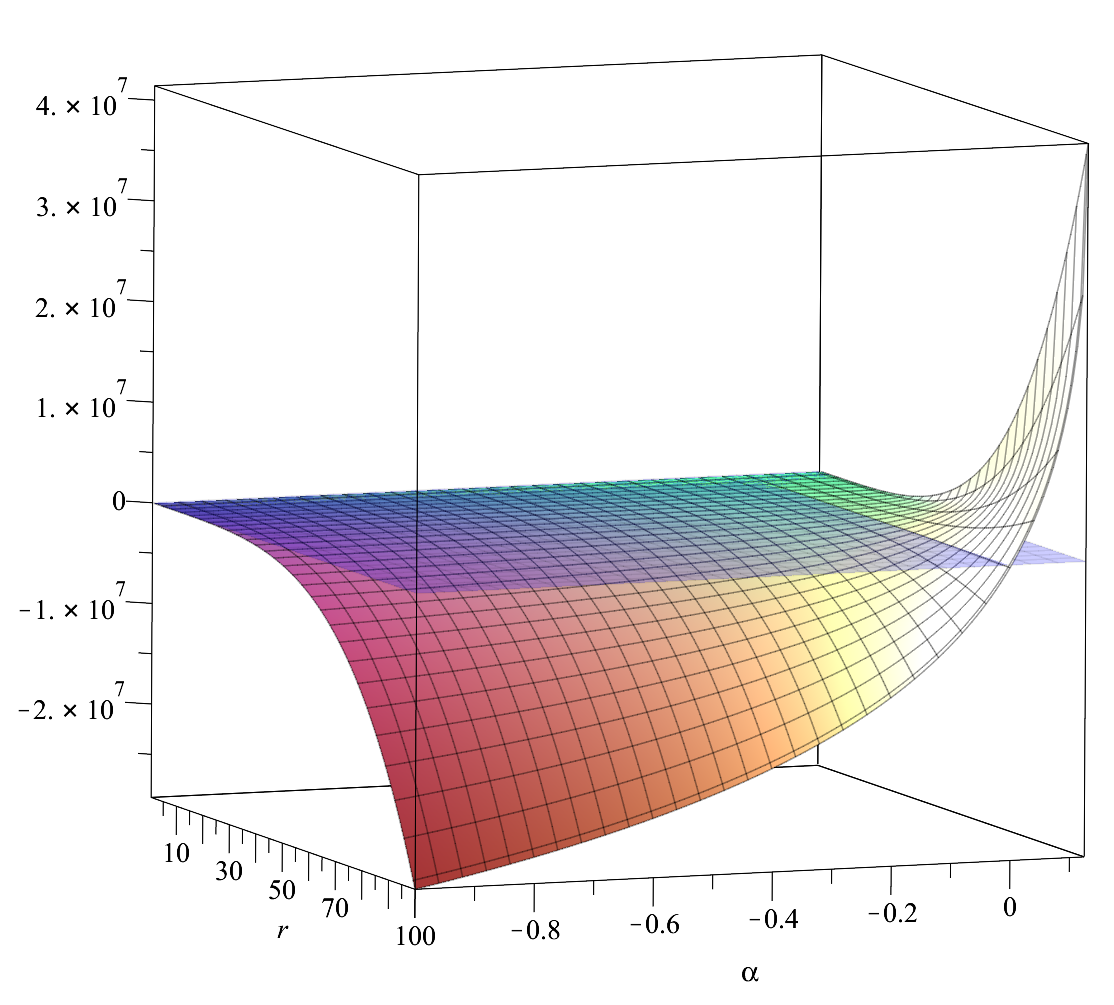}
	\caption{$\mathfrak{S}^\text{E}/(2\pi PL\Omega_{0,3})$ for a cubic toral EBG neutral black hole in 5-dimensions, with $L=1$ and $M=0.5$. The horizontal plane is the zero plane.}
	\label{fig}
\end{figure}

It is helpful to see visually the behavior of the function $\mathfrak{S}^\text{E}(r)$ as we vary $\alpha$, so let us consider the case $k=0$ in 5-dimensions and let $L=1$. Suppose the topology of the horizon is a cubic torus such that each copy of $S^1$ has periodicity $2\pi$. To ensure that the metric function Eq.(\ref{metricfn}) is real for all values of $M$, and in particular for the ground state without a black hole (i.e. $M=0$), we must have $\alpha < L^2/[4(D-3)]$. In our example, this means $\alpha < 1/8$. 
We show in Fig.(\ref{fig}) the behavior of $\mathfrak{S}^\text{E}$: it is negative for most values of $\alpha <0$, and positive for $\alpha \in (0, 1/8)$. (Actually, eikonal instability would rule out $\alpha > |1/8|$, at least in the planar limit \cite{1705.07732}.)

\begin{figure}[h!]
	\centering
	\includegraphics[scale=0.40]{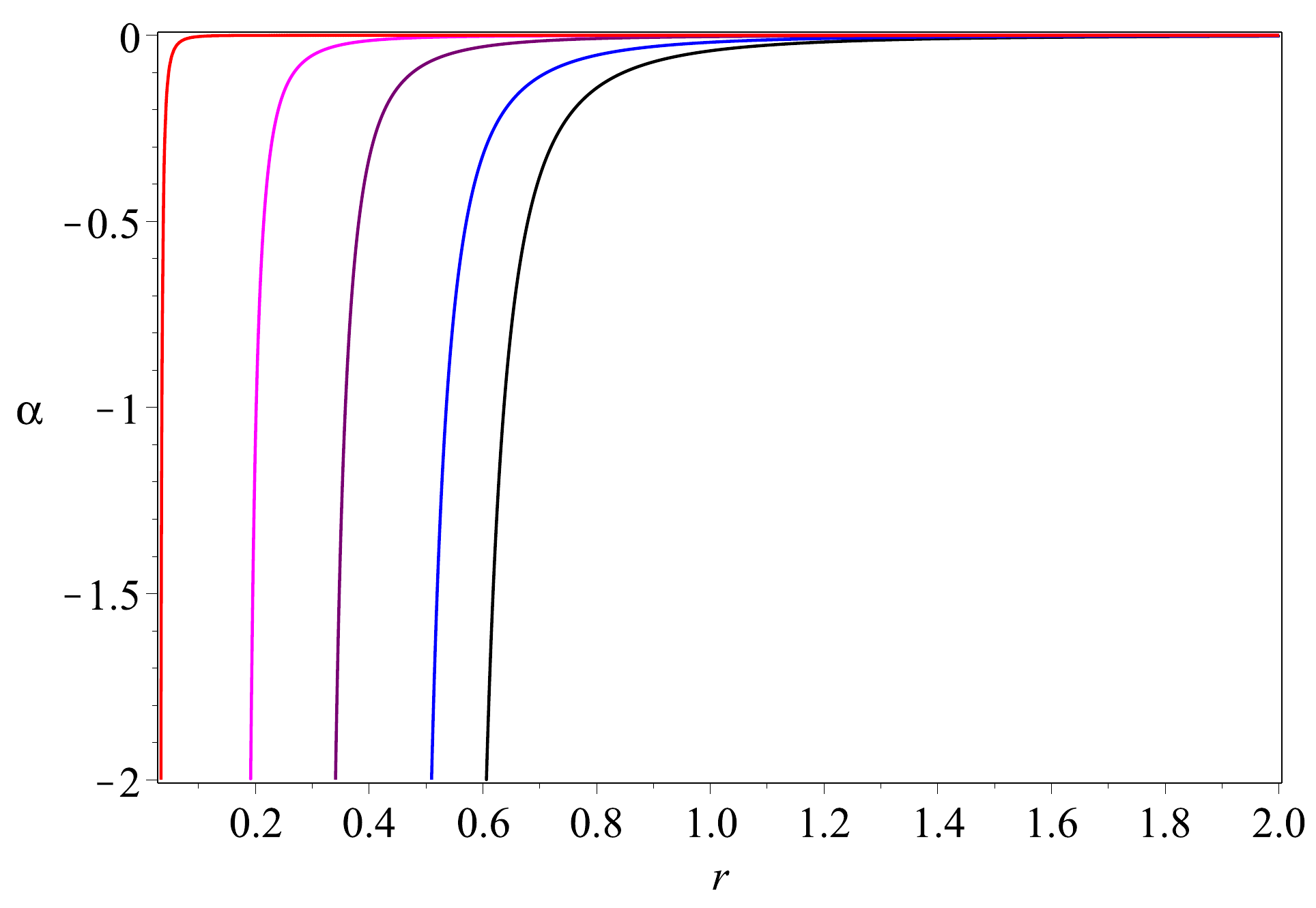}
	\caption{$\mathfrak{S}^\text{E}=0$ curve for a cubic toral EBG neutral black hole in 5-dimensions for $M=1,0.5,0.1,0.01,0.00001$ (right to left), where $L=1$.}
	\label{k0alpha}
\end{figure}

\begin{figure}[h!]
	\centering
	\includegraphics[scale=0.40]{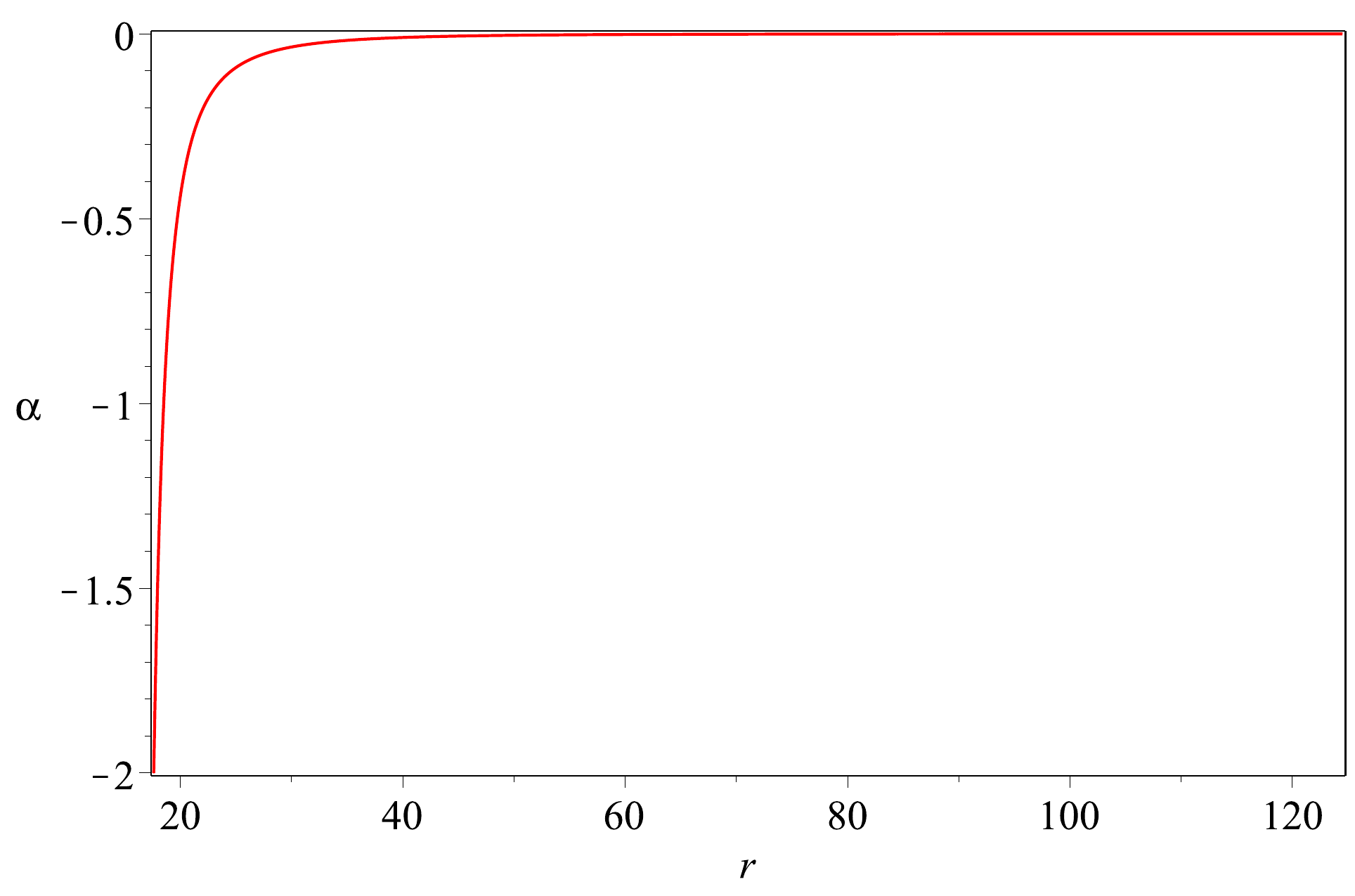}
	\caption{Brane action $\mathfrak{S}^\text{E}_\text{normalized}=0$ curve under the scaling symmetry $M \mapsto M/r_h^4$ and $r \mapsto r/r_h$. The horizon is now fixed at $r_h=3\pi^2/2$.}
	\label{renormalized}
\end{figure}

It turns out that the brane action is still positive for some negative values of $\alpha$ sufficiently close to the horizon. In Fig.(\ref{k0alpha}) we plot the curves along which $\mathfrak{S}^\text{E}(r)$ vanishes for various values of the mass. The region above a given curve corresponds to positive brane action, and likewise the region below it corresponds to negative brane action. We note that as $M$ increases, the curve is shifted towards the right, but so does the event horizon. If we denote the position where the action starts to become negative by $r_\text{neg}$, then $(r_\text{neg}-r_h)/r_h$ is constant (independent of the mass) for fixed $\alpha$. For $\alpha=-0.1$, this ratio is about $0.6617$. In fact, these curves exhibit a scaling symmetry under $M \mapsto M/r_h^4$ and $r \mapsto r/r_h$, which yields Fig.(\ref{renormalized}). 

The fact that the Euclidean brane action is negative already violates the consistency condition discussed in Sec.(\ref{II}) \cite{1504.07344}. 
But even if we only consider the Lorentzian picture, in which branes are nucleated in the region where the action is negative, we still have a problem. Causality dictates that the branes will back-react on the black hole geometry only after some time has passed. This is roughly the time it takes for a brane to free-fall towards the black hole (see, e.g., \cite{1201.6443}). The reason is that as branes are copiously produced in the bulk, if they are far away, they do not immediately affect the black hole in the bulk, whereas if they fall towards the black hole, then the event horizon, being teleological in nature, would start to expand outward in anticipation of the falling branes -- and thus the spacetime is no longer static.
Holography would be consistent if the field theory dual has a typical time scale larger than the instability time scale. This does not help us since the branes are nucleated close to the horizon, i.e., the instability time scale is short. 


The situation is actually worse: in the case \emph{without} a black hole ($M=0$), the brane action is negative for \emph{all} $\alpha < 0$. Thus \emph{the ground state is itself unstable}. Much of the above discussion also holds for the $k=1$ case, although the brane action remains positive near the origin $r=0$ when $M=0$. However, the ground state has no distinguished center, so $r=0$ is arbitrary. This implies that branes can be nucleated everywhere. The $k=-1$ case already suffers from brane production instability when the bulk gravity is GR (corresponds to the well known tachyonic instability of the boundary field scalar\footnote{``In this way, string theory reproduces
the instability that is evident in the field theory.'' -- Seiberg and Witten \cite{9903224}} \cite{9903224}), so we do not analyze it here. In any case, $k$ cannot affect the leading term of the EGB brane action at sufficiently large $r$, so the Euclidean brane action is still negative. 

\section{Discussion: Gauss-Bonnet Coupling Constant Is Positive}

Boulware and Deser already commented in 1985 \cite{BD} that a negative Gauss-Bonnet parameter $\alpha$ is pathological as it leads to a more singular behavior of the spacetime: a new curvature singularity arises at some nonzero value of $r$. However, since this singularity is behind the horizon (see also \cite{9706066}), one may still argue that this does not immediately rule out the possibility of $\alpha < 0$ EGB black holes (except perhaps for the reason raised in Footnote \ref{ft5}). 

In the bottom-up approach of holography, the $\alpha < 0$ case has also been considered in the literature. Often one hopes that in the low energy limit of string theory, stringy phenomena like effects of branes can be neglected. However, this is not always the case. This is reminiscent of the fact that some properties of the full theory can survive down to the infrared and imply constraints on the interactions that can be consistently added to the low energy EFT Lagrangian; see \cite{2106.08344} for an application in scalar-EGB theory. Holography consistency conditions impose constraints that render otherwise valid solutions (such as near extremal toral Reissner-Nordstr\"om black hole) in the anti-de Sitter bulk inconsistent \emph{even as effective low energy solutions}. This rules out the case for $\alpha < 0$, which is conjectured to be inconsistent by Sircar et al. in \cite{1602.07307} on phenomenological ground in holography.

Since $\alpha > 0$ is a consequence of inverse string tension being positive in the top-down approach, it is also a remarkable consistency check that holography requires \emph{precisely} this even if in the bottom-up approach one only takes EGB gravity as a modified theory of gravity and knows nothing about the relationship between $\alpha$ and string coupling in certain heterotic string theory embeddings. As discussed in \cite{0911.3160}, there are many different possibilities of realizing curvature squared corrections to Einstein-Hilbert action from string theory and M-theory \cite{9903210, 0904.4466}, so it is not obvious whether $\alpha < 0$ is consistent with \emph{some} version of string theory. However, if the holographic consistency condition in the form of ``isoperimetric inequality'' (\ref{brane}) -- a ``law of physics'' which has been checked in a myriad of highly nontrivial contexts and shown to hold in \cite{1504.07344}  --
is to be satisfied, then $\alpha < 0$ is indeed ruled out. 

Strictly speaking, our argument only shows that the usual static black hole solutions (along with their ground states) in EGB theory with $\alpha < 0$ are inconsistent, not the theory itself. However, taken together with the asymptotically flat case analysis of \cite{1608.02942}, the fact that the ground state itself is inconsistent strongly suggests that EGB gravity with $\alpha<0$ does indeed lie in the Swampland.

Nevertheless, we also need to be careful not to overstate the results. Just like the causality argument \cite{1407.5597} can only conclude that $\alpha$ has to be small in an effective EGB theory, but not when the full string theory is considered (with an infinite tower of higher spin particles), it is obvious that our analysis only concerns the validity of EGB gravity as an effective theory in the bulk. The results could change if one likewise considers other effects and other spin fields in the bulk, which in turn modify the black hole solutions and the ground states. 
Indeed, one might object that even as a low energy solution of string theory, the Regge slope $\alpha'$-correction (not to be confused with the GB parameter $\alpha$) needs to be taken into account especially since we are dealing with a higher curvature gravity theory \cite{9903210}. However, the point is that an $\alpha'$-corrected solution is no longer the original plain vanilla EGB solution discussed in this work. (Interestingly, the probe brane action to calculate the holographic consistency condition for the $\alpha'$-corrected Schwarzschild-AdS black hole in GR is unaffected by said correction \cite{1602.07177}, so it remains to be seen if and how the EGB case is different). 
Such possible loopholes would in principle allow a negative $\alpha$ in the bottom-up construction. Our work merely pointed out that the simplest EGB theory with $\alpha < 0$ is very likely to be problematic.

\begin{acknowledgments}
The author thanks the National Natural Science Foundation of China (Grant No.11922508) for funding support. He also thanks Brett McInnes for useful discussions, and colleagues (notably Ruihong Yue and Xiao-Mei Kuang) at the Center for Gravitation and Cosmology, Yangzhou University, for related discussions that partially inspired this work.
\end{acknowledgments}
\newpage



\begin{thebibliography}{99}

\bibitem{BD}
David G. Boulware, Stanley Deser, ``String-Generated Gravity Models'', {\hypersetup{urlcolor=vividviolet}\href{https://journals.aps.org/prl/abstract/10.1103/PhysRevLett.55.2656}{Phys. Rev. Lett. \textbf{55} (1985) 2656}}.

\bibitem{0911.3160}
Xian O. Camanho, Jose D. Edelstein, ``Causality Constraints in AdS/CFT From Conformal Collider Physics and Gauss-Bonnet Gravity'', {\hypersetup{urlcolor=vividviolet}\href{https://link.springer.com/article/10.1007\%2FJHEP04\%282010\%29007}{JHEP \textbf{04} (2010) 007}}, \href{https://arxiv.org/abs/0911.3160}{[arXiv:0911.3160 [hep-th]]}.


\bibitem{1905.03601}
Dra\v{z}en Glavan, Chunshan Lin, ``Einstein-Gauss-Bonnet Gravity in 4-Dimensional Spacetime'', {\hypersetup{urlcolor=vividviolet}\href{https://journals.aps.org/prl/abstract/10.1103/PhysRevLett.124.081301}{Phys. Rev. Lett. \textbf{124} (2020) 081301}}, \href{https://arxiv.org/abs/1905.03601}{[arXiv:1905.03601 [gr-qc]]}.

\bibitem{2003.11552}
Hong L\"u, Yi Pang, ``Horndeski Gravity as $D \rightarrow 4$ Limit of Gauss-Bonnet'', {\hypersetup{urlcolor=vividviolet}\href{https://www.sciencedirect.com/science/article/pii/S0370269320305207?via\%3Dihub}{Phys. Lett. B \textbf{809} (2020) 135717}}, \href{https://arxiv.org/abs/2003.11552}{[arXiv:2003.11552 [gr-qc]]}.

\bibitem{2004.02858}
Wen-Yuan Ai, ``A Note on the Novel 4D Einstein-Gauss-Bonnet Gravity'', {\hypersetup{urlcolor=vividviolet}\href{https://iopscience.iop.org/article/10.1088/1572-9494/aba242}{Commun. Theor. Phys. \textbf{72} (2020) 095402}}, \href{https://arxiv.org/abs/2004.02858}{[arXiv:2004.02858 [gr-qc]]}.

\bibitem{2004.03390}
Metin Gurses, Tahsin Cagri Sisman, Bayram Tekin, ``Is There a Novel Einstein-Gauss-Bonnet Theory in Four Dimensions?'', {\hypersetup{urlcolor=vividviolet}\href{https://link.springer.com/article/10.1140\%2Fepjc\%2Fs10052-020-8200-7}{Eur. Phys. J. C \textbf{80} (2020) 647}}, \href{https://arxiv.org/abs/2004.03390}{[arXiv:2004.03390 [gr-qc]]}.

\bibitem{2004.12998}
Julio Arrechea, Adri\`a Delhom, Alejandro Jim\'enez-Cano, ``Inconsistencies in Four-Dimensional Einstein-Gauss-Bonnet Gravity'', {\hypersetup{urlcolor=vividviolet}\href{https://iopscience.iop.org/article/10.1088/1674-1137/abc1d4}{Chin. Phys. C \textbf{45} (2021) 013107}},\href{https://arxiv.org/abs/2004.12998}{[arXiv:2004.12998 [gr-qc]]}.

\bibitem{2005.03859}
Katsuki Aoki, Mohammad Ali Gorji, Shinji Mukohyama, ``A Consistent Theory of $D \rightarrow 4$ Einstein-Gauss-Bonnet Gravity'', {\hypersetup{urlcolor=vividviolet}\href{https://www.sciencedirect.com/science/article/pii/S0370269320306468?via\%3Dihub}{Phys. Lett. B \textbf{810} (2020) 135843}}, \href{https://arxiv.org/abs/2005.03859}{[arXiv:2005.03859 [gr-qc]]}.

\bibitem{2005.12292}
Damien A. Easson, Tucker Manton, Andrew Svesko, ``$D \rightarrow 4$ Einstein-Gauss-Bonnet Gravity and Beyond'', {\hypersetup{urlcolor=vividviolet}\href{https://iopscience.iop.org/article/10.1088/1475-7516/2020/10/026}{JCAP \textbf{10} (2020) 026}}, \href{https://arxiv.org/abs/2005.12292}{[arXiv:2005.12292 [hep-th]]}.

\bibitem{2009.10715}
Julio Arrechea, Adri\`{a} Delhom, Alejandro Jim\'{e}nez-Cano, ``Comment on `Einstein-Gauss-Bonnet Gravity in Four-Dimensional Spacetime''', {\hypersetup{urlcolor=vividviolet}\href{https://arxiv.org/abs/2009.10715}{Phys. Rev. Lett. \textbf{125} (2020) 149002}}, \href{https://arxiv.org/abs/2009.10715}{[arXiv:2009.10715 [gr-qc]]}.


\bibitem{Z}
Barton Zwiebach, ``Curvature Squared Terms and String Theories'', {\hypersetup{urlcolor=vividviolet}\href{https://www.sciencedirect.com/science/article/abs/pii/0370269385916168?via\%3Dihub}{Phys. Lett. B \textbf{156} (1985) 315}}.

\bibitem{GW}
David J. Gross, Edward Witten, ``Superstring Modifications of Einstein's Equations'', {\hypersetup{urlcolor=vividviolet}\href{https://www.sciencedirect.com/science/article/abs/pii/0550321386904293}{Nucl. Phys. B \textbf{277} (1986) 1}}.

\bibitem{GS}
David J. Gross, John H. Sloan, ``The Quartic Effective action for the Heterotic String'', {\hypersetup{urlcolor=vividviolet}\href{https://www.sciencedirect.com/science/article/abs/pii/0550321387904652?via\%3Dihub}{Nucl. Phys. B \textbf{291} (1987) 41}}.

\bibitem{cai}
Rong-Gen Cai, ``Gauss-Bonnet Black Holes in AdS Spaces'',  {\hypersetup{urlcolor=vividviolet}\href{https://journals.aps.org/prd/abstract/10.1103/PhysRevD.65.084014}{Phys. Rev. D \textbf{65} (2002) 084014}}, \href{https://arxiv.org/abs/hep-th/0109133}{[arXiv:hep-th/0109133]}.

\bibitem{0504127}
Takashi Torii, Hideki Maeda, ``Spacetime Structure of Static Solutions in Gauss-Bonnet Gravity: Neutral Case'', {\hypersetup{urlcolor=vividviolet}\href{https://journals.aps.org/prd/abstract/10.1103/PhysRevD.71.124002}{Phys. Rev. D \textbf{71} (2005) 124002}}, \href{https://arxiv.org/abs/hep-th/0504127}{[arXiv:hep-th/0504127]}.

\bibitem{0504141}
Takashi Torii, Hideki Maeda,  ``Spacetime Structure of Static Solutions in Gauss-Bonnet Gravity: Charged Case'', {\hypersetup{urlcolor=vividviolet}\href{https://journals.aps.org/prd/abstract/10.1103/PhysRevD.72.064007}{Phys. Rev. D \textbf{72} (2005) 064007}}, \href{https://arxiv.org/abs/hep-th/0504141}{[arXiv:hep-th/0504141]}.


\bibitem{1807.00591}
Constantin Bachas, Ioannis Lavdas, ``Massive Anti-de Sitter Gravity from String Theory'', {\hypersetup{urlcolor=vividviolet}\href{https://link.springer.com/article/10.1007/JHEP11(2018)003}{JHEP \textbf{11} (2018) 003}}, \href{https://arxiv.org/abs/1807.00591}{[arXiv:1807.00591 [hep-th]]}.

\bibitem{2106.04614}
Dieter Lust, Chrysoula Markou, Pouria Mazloumi, Stephan Stieberger, ``Extracting Bigravity from String Theory'', {\hypersetup{urlcolor=vividviolet}\href{https://link.springer.com/article/10.1007\%2FJHEP12\%282021\%29220}{JHEP \textbf{12} (2021) 220}}, \href{https://arxiv.org/abs/2106.04614}{[arXiv:2106.04614 [hep-th]]}.


\bibitem{1211.0010}
Stefan Janiszewski, Andreas Karch, ``String Theory Embeddings of Non-Relativistic Field Theories and Their Holographic Ho\v{r}ava Gravity Duals'', {\hypersetup{urlcolor=vividviolet}\href{https://journals.aps.org/prl/abstract/10.1103/PhysRevLett.110.081601}{Phys. Rev. Lett. \textbf{110} (2013) 8, 081601}}, \href{https://arxiv.org/abs/1211.0010}{[arXiv:1211.0010 [hep-th]]}.

\bibitem{1512.03554}
Marika Taylor, ``Lifshitz Holography'', {\hypersetup{urlcolor=vividviolet}\href{https://iopscience.iop.org/article/10.1088/0264-9381/33/3/033001}{Class. Quant. Grav. \textbf{33} (2016)  033001}}, \href{https://arxiv.org/abs/1512.03554}{[arXiv:1512.03554 [hep-th]]}.

\bibitem{0605264}
Hirosi Ooguri, Cumrun Vafa, ``On the Geometry of the String Landscape and the Swampland'', {\hypersetup{urlcolor=vividviolet}\href{https://linkinghub.elsevier.com/retrieve/pii/S0550321306008455}{Nucl. Phys. B \textbf{766} (2007) 21}}, \href{https://arxiv.org/abs/hep-th/0605264}{[arXiv:hep-th/0605264]}.

\bibitem{1903.06239}
Eran Palti, ``The Swampland: Introduction and Review'', {\hypersetup{urlcolor=vividviolet}\href{https://onlinelibrary.wiley.com/doi/abs/10.1002/prop.201900037}{Fortsch. Phys. \textbf{67} (2019) 6, 1900037}}, \href{https://arxiv.org/abs/1903.06239}{[arXiv:1903.06239 [hep-th]]}.

\bibitem{1406.0677}
Keisuke Izumi, ``Causal Structures in Gauss-Bonnet Gravity'', {\hypersetup{urlcolor=vividviolet}\href{https://journals.aps.org/prd/abstract/10.1103/PhysRevD.90.044037}{Phys. Rev. D \textbf{90} (2014) 044037}}, \href{https://arxiv.org/abs/1406.0677}{[arXiv:1406.0677 [gr-qc]]}.

\bibitem{1406.3379}
Harvey S. Reall, Norihiro Tanahashi, Benson Way, ``Causality and Hyperbolicity of Lovelock Theories'',  {\hypersetup{urlcolor=vividviolet}\href{https://iopscience.iop.org/article/10.1088/0264-9381/31/20/205005}{Class. Quant. Grav. \textbf{31} (2014) 205005}}, \href{https://arxiv.org/abs/1406.3379}{[arXiv:1406.3379 [hep-th]]}.

\bibitem{1407.5597}
Xian O. Camanho, Jose D. Edelstein, Juan Maldacena, Alexander Zhiboedov, ``Causality Constraints on Corrections to the Graviton Three-Point Coupling'', {\hypersetup{urlcolor=vividviolet}\href{https://link.springer.com/article/10.1007\%2FJHEP02\%282016\%29020}{JHEP \textbf{02} (2016) 020}}, \href{https://arxiv.org/abs/1407.5597}{[arXiv:1407.5597 [hep-th]]}.

\bibitem{1508.05303}
Giuseppe Papallo, Harvey S. Reall, ``Graviton Time Delay and a Speed Limit for Small Black Holes in Einstein-Gauss-Bonnet Theory'', {\hypersetup{urlcolor=vividviolet}\href{https://link.springer.com/article/10.1007\%2FJHEP11\%282015\%29109}{JHEP \textbf{11} (2015) 109}}, \href{https://arxiv.org/abs/1508.05303}{[arXiv:1508.05303 [gr-qc]]}.

\bibitem{2101.02461}
Li-Ming Cao, Liang-Bi Wu, ``Hyperbolicity and Causality of Einstein-Gauss-Bonnet Gravity in Warped Product Spacetimes'', {\hypersetup{urlcolor=vividviolet}\href{https://journals.aps.org/prd/abstract/10.1103/PhysRevD.103.064054}{Phys. Rev. D \textbf{103} (2021) 064054}}, \href{https://arxiv.org/abs/2101.02461}{[arXiv:2101.02461 [gr-qc]]}.

\bibitem{1401.5089}
Shamik Banerjee, Arpan Bhattacharyya, Apratim Kaviraj, Kallol Sen, Aninda Sinha, ``Constraining Gravity Using Entanglement in AdS/CFT'' , {\hypersetup{urlcolor=vividviolet}\href{https://link.springer.com/article/10.1007\%2FJHEP05\%282014\%29029}{JHEP \textbf{05} (2014) 029}}, \href{https://arxiv.org/abs/1401.5089}{[arXiv:1401.5089 [hep-th]]}.

\bibitem{1610.06078}
Tomas Andrade, Elena Caceres, Cynthia Keeler, ``Boundary Causality vs Hyperbolicity for Spherical Black Holes in Gauss-Bonnet'', {\hypersetup{urlcolor=vividviolet}\href{https://iopscience.iop.org/article/10.1088/1361-6382/aa7101}{Class. Quant. Grav. \textbf{34} (2017) 135003}}, \href{https://arxiv.org/abs/1610.06078}{[arXiv:1610.06078 [hep-th]]}.

\bibitem{0802.3318}
Mauro Brigante, Hong Liu, Robert C. Myers, Stephen Shenker, Sho Yaida, ``Viscosity Bound and Causality Violation'', {\hypersetup{urlcolor=vividviolet}\href{https://journals.aps.org/prl/abstract/10.1103/PhysRevLett.100.191601}{Phys. Rev. Lett. \textbf{100} (2008) 191601}}, \href{https://arxiv.org/abs/0802.3318}{[arXiv:0802.3318 [hep-th]]}.

\bibitem{0906.2922}
Alex Buchel, Robert C. Myers, ``Causality of Holographic Hydrodynamics'', {\hypersetup{urlcolor=vividviolet}\href{https://iopscience.iop.org/article/10.1088/1126-6708/2009/08/016}{JHEP \textbf{08} (2009) 016}}, \href{https://arxiv.org/abs/0906.2922}{[arXiv:0906.2922 [hep-th]]}.

\bibitem{0911.4257}
Alex Buchel, Jorge Escobedo, Robert C. Myers, Miguel F. Paulos, Aninda Sinha, Michael Smolkin, ``Holographic GB Gravity in Arbitrary Dimensions'', {\hypersetup{urlcolor=vividviolet}\href{https://link.springer.com/article/10.1007\%2FJHEP03\%282010\%29111}{JHEP \textbf{03} (2010) 111}}, \href{https://arxiv.org/abs/0911.4257}{[arXiv:0911.4257 [hep-th]]}.




\bibitem{0007021}
Sijie Gao, Robert M. Wald, ``Theorems on Gravitational Time Delay and Related Issues'', {\hypersetup{urlcolor=vividviolet}\href{https://iopscience.iop.org/article/10.1088/0264-9381/17/24/305}{Class. Quant. Grav. \textbf{17} (2000) 4999}}, \href{https://arxiv.org/abs/gr-qc/0007021}{[arXiv:gr-qc/0007021]}.

\bibitem{2004.12191}
Xian-Hui Ge, Sang-Jin Sin, ``Causality of Black Holes in 4-Dimensional Einstein-Gauss-Bonnet-Maxwell Theory'',  {\hypersetup{urlcolor=vividviolet}\href{https://link.springer.com/article/10.1140\%2Fepjc\%2Fs10052-020-8288-9}{Eur. Phys. J. C \textbf{80} (2020) 695}}, \href{https://arxiv.org/abs/2004.12191}{[arXiv:2004.12191 [hep-th]]}.

\bibitem{2006.15017}
Timothy Clifton, Pedro Carrilho, Pedro G. S. Fernandes, David J. Mulryne, ``Observational Constraints on the Regularized 4D Einstein-Gauss-Bonnet Theory of Gravity'', {\hypersetup{urlcolor=vividviolet}\href{https://journals.aps.org/prd/abstract/10.1103/PhysRevD.102.084005}{Phys. Rev. D \textbf{102} (2020) 084005}}, \href{https://arxiv.org/abs/2006.15017}{[arXiv:2006.15017 [gr-qc]]}.


\bibitem{1604.03944}
Netta Engelhardt, Sebastian Fischetti, ``The Gravity Dual of Boundary Causality'', {\hypersetup{urlcolor=vividviolet}\href{https://iopscience.iop.org/article/10.1088/0264-9381/33/17/175004}{Class. Quant. Grav. \textbf{33} (2016) 175004}}, \href{https://arxiv.org/abs/1604.03944}{[arXiv:1604.03944 [hep-th]]}.

\bibitem{1901.03928}
Edward Witten, ``Light Rays, Singularities, and All That'', {\hypersetup{urlcolor=vividviolet}\href{https://journals.aps.org/rmp/abstract/10.1103/RevModPhys.92.045004}{Rev. Mod. Phys. \textbf{92} (2000) 45004}}, \href{https://arxiv.org/abs/1901.03928}{[arXiv:1901.03928 [hep-th]]}.

\bibitem{0104066}
Giuseppe Policastro, Dan T. Son, Andrei O. Starinets, ``The Shear Viscosity of Strongly Coupled $\mathcal{N}=4$ Supersymmetric Yang-Mills Plasma'', {\hypersetup{urlcolor=vividviolet}\href{https://journals.aps.org/prl/abstract/10.1103/PhysRevLett.87.081601}{Phys. Rev. Lett. \textbf{87} (2001) 081601}}, \href{https://arxiv.org/abs/hep-th/0104066}{[arXiv:hep-th/0104066]}.

\bibitem{0309213}
Pavel Kovtun, Dam T. Son, Andrei O. Starinets, ``Holography and Hydrodynamics: Diffusion on Stretched Horizons'',  {\hypersetup{urlcolor=vividviolet}\href{https://iopscience.iop.org/article/10.1088/1126-6708/2003/10/064}{JHEP \textbf{10} (2003) 064}}, \href{https://arxiv.org/abs/hep-th/0309213}{[arXiv:hep-th/0309213]}.

\bibitem{0712.0743}
Yevgeny Kats, Pavel Petrov, ``Effect of Curvature Squared Corrections in AdS on the Viscosity of the Dual Gauge Theory'', {\hypersetup{urlcolor=vividviolet}\href{https://iopscience.iop.org/article/10.1088/1126-6708/2009/01/044}{JHEP \textbf{01} (2009) 044}}, \href{https://arxiv.org/abs/0712.0743}{[arXiv:0712.0743 [hep-th]]}.


\bibitem{0812.2521}
Alex Buchel, Robert C. Myers, Aninda Sinha, ``Beyond $\eta/s = 1/4\pi$'', 	{\hypersetup{urlcolor=vividviolet}\href{https://iopscience.iop.org/article/10.1088/1126-6708/2009/03/084}{JHEP \textbf{03} (2009) 084}}, \href{https://arxiv.org/abs/0812.2521}{[arXiv:0812.2521 [hep-th]]}.


\bibitem{0712.0805}
Mauro Brigante, Hong Liu, Robert C. Myers, Stephen Shenker, Sho Yaida, ``Viscosity Bound Violation in Higher Derivative Gravity'', {\hypersetup{urlcolor=vividviolet}\href{https://journals.aps.org/prd/abstract/10.1103/PhysRevD.77.126006}{Phys. Rev. D \textbf{77} (2008) 126006}}, \href{https://arxiv.org/abs/0712.0805}{[arXiv:0712.0805 [hep-th]]}.


\bibitem{0808.1919}
Ishwaree P. Neupane, Naresh Dadhich, ``Entropy Bound and Causality Violation in Higher Curvature Gravity'', {\hypersetup{urlcolor=vividviolet}\href{https://iopscience.iop.org/article/10.1088/0264-9381/26/1/015013}{Class. Quant. Grav. \textbf{26} (2009) 015013}}, \href{https://arxiv.org/abs/0808.1919}{[arXiv:0808.1919 [hep-th]]}.

\bibitem{0904.4805}
Ishwaree P Neupane, ``Black Holes, Entropy Bound and Causality Violation'',  {\hypersetup{urlcolor=vividviolet}\href{https://www.worldscientific.com/doi/abs/10.1142/S0217751X09047235}{Int. J. Mod. Phys. A \textbf{24} (2009) 3584}}, \href{https://arxiv.org/abs/0904.4805}{[arXiv:0904.4805 [gr-qc]]}.

\bibitem{1701.01652}
Roman A. Konoplya, Alexander Zhidenko, ``Eikonal Instability of Gauss-Bonnet–(Anti-)–de Sitter Black Holes'', {\hypersetup{urlcolor=vividviolet}\href{https://journals.aps.org/prd/abstract/10.1103/PhysRevD.95.104005}{Phys. Rev. D \textbf{95} (2017) 104005}}, \href{https://arxiv.org/abs/1701.01652v6}{[arXiv:1701.01652 [hep-th]]}.

\bibitem{1705.07732}
Roman A. Konoplya, Alexander Zhidenko, ``Quasinormal Modes of Gauss-Bonnet-AdS Black Holes: Towards Holographic Description of Finite Coupling'', {\hypersetup{urlcolor=vividviolet}\href{https://link.springer.com/article/10.1007\%2FJHEP09\%282017\%29139}{JHEP \textbf{09} (2017) 139}}, \href{https://arxiv.org/abs/1705.07732v2}{[arXiv:1705.07732 [hep-th]]}.

\bibitem{1103.3982}
Jian-Pin Wu, ``Holographic Fermions In Charged Gauss-Bonnet Black Hole'', {\hypersetup{urlcolor=vividviolet}\href{https://link.springer.com/article/10.1007\%2FJHEP07\%282011\%29106}{JHEP \textbf{07} (2011) 106}}, \href{https://arxiv.org/abs/1103.3982}{[arXiv:1103.3982 [hep-th]]}.

\bibitem{1305.4841}
Xiao-Xiong Zeng, Wen-Biao Liu, ``Holographic Thermalization in Gauss-Bonnet Gravity'', {\hypersetup{urlcolor=vividviolet}\hypersetup{urlcolor=vividviolet}\href{https://linkinghub.elsevier.com/retrieve/pii/S0370269313006886}{Phys. Lett. B \textbf{726} (2013) 481}}, \href{https://arxiv.org/abs/1305.4841}{[arXiv:1305.4841 [hep-th]]}.

\bibitem{1343004}
Sa\v{s}o Grozdanov, Andrei Olegovich Starinets, ``Zero-Viscosity Limit in a Holographic Gauss–Bonnet Liquid'', {\hypersetup{urlcolor=vividviolet}\href{https://link.springer.com/article/10.1007\%2Fs11232-015-0245-7}{Theor. Math. Phys. \textbf{182} (2015) 61}}, Teor. Mat. Fiz. \textbf{182} (2014) 76.

\bibitem{1311.3053}
Wei Xu, Hao Xu, Liu Zhao, ``Gauss-Bonnet Coupling Constant as a Free Thermodynamical Variable and the Associated Criticality'', {\hypersetup{urlcolor=vividviolet}\href{https://link.springer.com/article/10.1140\%2Fepjc\%2Fs10052-014-2970-8}{Eur. Phys. J. C \textbf{74} (2014) 2970}}, \href{https://arxiv.org/abs/1311.3053}{[arXiv:1311.3053 [gr-qc]]}.

\bibitem{1508.03364}
Shao-Jun Zhang, Elcio Abdalla, ``Holographic Schwinger Effect In a Confining Background With Gauss-Bonnet Corrections'', {\hypersetup{urlcolor=vividviolet}\href{https://link.springer.com/article/10.1007\%2Fs10714-016-2056-z}{Gen. Rel. Grav. \textbf{48} (2016) 5, 60}}, \href{https://arxiv.org/abs/1508.03364v3}{[arXiv:1508.03364 [hep-th]]}.

\bibitem{1608.04208}
Song He, Li-Fang Li, Xiao-Xiong Zeng, ``Holographic Van der Waals-Like Phase Transition in the Gauss-Bonnet Gravity'', {\hypersetup{urlcolor=vividviolet}\href{https://www.sciencedirect.com/science/article/pii/S0550321316303911?via\%3Dihub}{Nucl. Phys. B \textbf{915} (2017) 243}}, \href{https://arxiv.org/abs/1608.04208}{[arXiv:1608.04208 [hep-th]]}. 

\bibitem{1712.02772}
Jorge Casalderrey-Solana, Nikola I. Gushterov, Ben Meiring, ``Resurgence and Hydrodynamic Attractors in Gauss-Bonnet Holography'', {\hypersetup{urlcolor=vividviolet}\href{https://dx.doi.org/10.1007/JHEP04\%282018\%29042}{JHEP \textbf{04} (2018) 042}}, \href{https://arxiv.org/abs/1712.02772}{[arXiv:1712.02772 [hep-th]]}.

\bibitem{Lu}
Jun-Wang Lu, Ya-Bo Wu, Hai-Min Liu, Yin-Shuan Ren, Mo-Lin Liu, ``Holographic Vector Superconductor in Gauss–Bonnet Gravity'', {\hypersetup{urlcolor=vividviolet}\href{https://www.sciencedirect.com/science/article/pii/S0550321316000110}{Nucl. Phys. B \textbf{903} (2016) 360}}.

\bibitem{2101.00882}
Cao H. Nam, ``A More Realistic Holographic Model of Color Superconductivity in Einstein-Gauss-Bonnet Gravity'', {\hypersetup{urlcolor=vividviolet}\href{https://journals.aps.org/prd/abstract/10.1103/PhysRevD.104.046006}{Phys. Rev. D \textbf{104} (2021) 046006}}, \href{https://arxiv.org/abs/2101.00882}{[arXiv:2101.00882 [hep-th]]}.


\bibitem{0407083}
Sachiko Ogushi, Misao Sasaki, ``Holography in Einstein Gauss-Bonnet Gravity'', {\hypersetup{urlcolor=vividviolet}\href{https://academic.oup.com/ptp/article/113/5/979/1821494}{Prog. Theor. Phys. \textbf{113} (2005) 979}}, \href{https://arxiv.org/abs/hep-th/0407083}{[arXiv:hep-th/0407083]}.

\bibitem{1610.08987}
Tomas Andrade, Jorge Casalderrey-Solana, Andrej Ficnar, ``Holographic Isotropisation in Gauss-Bonnet Gravity'', {\hypersetup{urlcolor=vividviolet}\href{https://link.springer.com/article/10.1007\%2FJHEP02\%282017\%29016}{JHEP \textbf{02} (2017) 016}}, \href{https://arxiv.org/abs/1610.08987}{[arXiv:1610.08987 [hep-th]]}. 

\bibitem{2102.12171}
Yong-Zhuang Li, Cheng-Yong Zhang, Xiao-Mei Kuang, ``Entanglement Wedge Cross Section with Gauss-Bonnet Corrections and Thermal Quench'', 	{\hypersetup{urlcolor=vividviolet}\href{https://link.springer.com/article/10.1007\%2Fs11433-021-1791-1}{Sci. China-Phys. Mech. Astron. \textbf{64} (2021) 120413}}, \href{https://arxiv.org/abs/2102.12171}{[arXiv:2102.12171 [hep-th]]}.

\bibitem{1512.05666}
Elena Caceres, Manuel Sanchez, Julio Virrueta, ``Holographic Entanglement Entropy in Time Dependent Gauss-Bonnet Gravity'', {\hypersetup{urlcolor=vividviolet}\href{https://link.springer.com/article/10.1007/JHEP09(2017)127}{JHEP \textbf{09} (2017) 127}}, \href{https://arxiv.org/abs/1512.05666}{[arXiv:1512.05666 [hep-th]]}.

\bibitem{2106.13942}
Yuan-zhang Cui, Wei Xu, Bin Zhu, ``Hawking-Page Transition With Reentrance and Triple Point in Gauss-Bonnet Gravity'', \href{https://arxiv.org/abs/2106.13942}{[arXiv:2106.13942 [gr-qc]]}.

\bibitem{1602.07307}
Nilanjan Sircar, Jacob Sonnenschein, Walter Tangarife, ``Extending the Scope of Holographic Mutual Information and Chaotic Behavior'', {\hypersetup{urlcolor=vividviolet}\href{https://link.springer.com/article/10.1007\%2FJHEP05\%282016\%29091}{JHEP \textbf{05} (2016) 091}}, \href{https://arxiv.org/abs/1602.07307}{[arXiv:1602.07307 [hep-th]]}.

\bibitem{2103.00257}
Chen-Hao Wu, Ya-Peng Hu, Hao Xu, ``Hawking Evaporation of Einstein-Gauss-Bonnet AdS Black Holes In $D \geqslant 4$ Dimensions'', {\hypersetup{urlcolor=vividviolet}\href{https://link.springer.com/article/10.1140\%2Fepjc\%2Fs10052-021-09140-6}{Eur. Phys. J. C \textbf{81} (2021) 351}}, \href{https://arxiv.org/abs/2103.00257}{[arXiv:2103.00257 [hep-th]]}.

\bibitem{1608.02942}
Clifford Cheung, Grant N. Remmen, ``Positivity of Curvature-Squared Corrections in Gravity'', {\hypersetup{urlcolor=vividviolet}\href{https://journals.aps.org/prl/abstract/10.1103/PhysRevLett.118.051601}{Phys. Rev. Lett. \textbf{18} (2017) 051601}}, \href{https://arxiv.org/abs/1608.02942}{[arXiv:1608.02942 [hep-th]]}.

\bibitem{1501.00007}
Veronika E. Hubeny, ``The AdS/CFT Correspondence'', {\hypersetup{urlcolor=vividviolet}\href{https://doi.org/10.1088/0264-9381/32/12/124010}{Class. Quant. Grav. \textbf{32} (2015) 124010}}, \href{https://arxiv.org/abs/1501.00007}{[arXiv:1501.00007 [gr-qc]]}.

\bibitem{1409.3575}
Makoto Natsuume, ``AdS/CFT Duality User Guide'', {\hypersetup{urlcolor=vividviolet}\href{https://link.springer.com/book/10.1007\%2F978-4-431-55441-7}{Lect. Notes Phys. \textbf{903} (2015) 1}}, \href{https://arxiv.org/abs/1409.3575}{[arXiv:1409.3575 [hep-th]}.

\bibitem{1310.6788}
Frank Ferrari, ``Gauge Theories, D-Branes and Holography'', {\hypersetup{urlcolor=vividviolet}\href{https://doi.org/10.1016/j.nuclphysb.2014.01.007}{Nucl. Phys. B \textbf{80} (2014) 247}}, \href{https://arxiv.org/abs/1310.6788}{[arXiv:1310.6788 [hep-th]]}.

\bibitem{1311.4520}
Frank Ferrari, ``D-Brane Probes in the Matrix Model'', {\hypersetup{urlcolor=vividviolet}\href{https://linkinghub.elsevier.com/retrieve/pii/S0550321313006068}{Nucl. Phys. B \textbf{880} (2014) 290}}, \href{https://arxiv.org/abs/1311.4520}{[arXiv:1311.4520 [hep-th]]}.

\bibitem{1411.1887}
Frank Ferrari, Antonin Rovai, ``Holography, Probe Branes and Isoperimetric Inequalities'', {\hypersetup{urlcolor=vividviolet}\href{https://www.sciencedirect.com/science/article/pii/S0370269315004268?via\%3Dihub}{Phys. Lett. B \textbf{747} (2015) 212}}, \href{https://arxiv.org/abs/1411.1887}{[arXiv:1411.1887 [hep-th]]}.

\bibitem{1602.07177}
Frank Ferrari, Antonin Rovai, ``Gravity and On-Shell Probe Actions'', {\hypersetup{urlcolor=vividviolet}\href{https://link.springer.com/article/10.1007\%2FJHEP08\%282016\%29047}{JHEP \textbf{08} (2016) 047}}, \href{https://arxiv.org/abs/1602.07177}{[arXiv:1602.07177 [hep-th]]}.

\bibitem{1610.07313}
Brett McInnes, ``Isoperimetric Inequalities and Magnetic Fields at CERN'', Asia-Pacific Mathematics Newsletter \textbf{6} (2016) 10, \href{https://arxiv.org/abs/1610.07313}{[arXiv:1610.07313 [hep-th]]}.

\bibitem{Wang1}
Xiaodong Wang, ``On Conformally Compact Einstein Manifolds'', {\hypersetup{urlcolor=vividviolet}\href{https://www.intlpress.com/site/pub/pages/journals/items/mrl/content/vols/0008/0005/a009/index.php}{Math. Res. Lett. \textbf{8} (2001) 671}}.

\bibitem{Wang2}
Xiaodong Wang, ``A New Proof of Lee’s Theorem on the Spectrum of Conformally Compact Einstein Manifolds'', {\hypersetup{urlcolor=vividviolet}\href{https://dx.doi.org/10.4310/CAG.2002.v10.n3.a7}{Comm. Anal. Geom. \textbf{10} (2002) 647}}.

\bibitem{1504.07344}
Brett McInnes, Yen Chin Ong, ``When Is Holography Consistent?'', {\hypersetup{urlcolor=vividviolet}\href{https://linkinghub.elsevier.com/retrieve/pii/S0550321315002412}{Nucl. Phys. B \textbf{898} (2015) 197}}, \href{https://arxiv.org/abs/1504.07344}{[arXiv:1504.07344 [hep-th]]}.

\bibitem{WY}
Edward Witten, Shing-Tung Yau, ``Connectedness of the Boundary in the AdS/CFT Correspondence'', {\hypersetup{urlcolor=vividviolet}\href{https://www.intlpress.com/site/pub/pages/journals/items/atmp/content/vols/0003/0006/a001/}{Adv. Theor. Math. Phys. \textbf{3} (1999) 1635}}, \href{https://arxiv.org/abs/hep-th/9910245}{[arXiv:hep-th/9910245]}.

\bibitem{CG}
Mingliang Cai, Gregory J. Galloway, ``Boundaries of Zero Scalar Curvature in the AdS/CFT Correspondence'', {\hypersetup{urlcolor=vividviolet}\href{https://www.intlpress.com/site/pub/pages/journals/items/atmp/content/vols/0003/0006/a004/}{Adv. Theor. Math. Phys. \textbf{3} (1999) 1769}}, \href{https://arxiv.org/abs/hep-th/0003046}{[arXiv:hepth/0003046]}.

\bibitem{1005.4439}
Jos\'e L. F. Barb\'on, Javier Mart\'inez-Mag\'an, ``Spontaneous Fragmentation of Topological Black Holes'', {\hypersetup{urlcolor=vividviolet}\href{https://link.springer.com/article/10.1007\%2FJHEP08\%282010\%29031}{JHEP \textbf{08} (2010) 031}}, \href{https://arxiv.org/abs/1005.4439}{[arXiv:1005.4439 [hep-th]]}.

\bibitem{2005.12075}
Yen Chin Ong, ``Schwinger Pair Production and the Extended Uncertainty Principle: Can Heuristic Derivations Be Trusted?'', {\hypersetup{urlcolor=vividviolet}\href{https://link.springer.com/article/10.1140\%2Fepjc\%2Fs10052-020-8363-2}{Eur. Phys. J. C \textbf{80} (2020) 8, 777}}, \href{https://arxiv.org/abs/2005.12075}{[arXiv:2005.12075 [gr-qc]]}.

\bibitem{9903224}
Nathan Seiberg, Edward Witten, ``The D1/D5 System And Singular CFT'', {\hypersetup{urlcolor=vividviolet}\href{https://iopscience.iop.org/article/10.1088/1126-6708/1999/04/017}{JHEP \textbf{04} (1999) 017}}, \href{https://arxiv.org/abs/hep-th/9903224}{[arXiv:hep-th/9903224]}.

\bibitem{0409242}
Matthew Kleban, Massimo Porrati, Raul Rabadan, ``Stability in Asymptotically AdS Spaces'', {\hypersetup{urlcolor=vividviolet}\href{https://iopscience.iop.org/article/10.1088/1126-6708/2005/08/016}{JHEP \textbf{08} (2005) 016}}, \href{https://arxiv.org/abs/hep-th/0409242}{[arXiv:hep-th/0409242]}.

\bibitem{0401024}
Juan Maldacena, Liat Maoz, ``Wormholes in AdS'', {\hypersetup{urlcolor=vividviolet}\href{https://iopscience.iop.org/article/10.1088/1126-6708/2004/02/053}{JHEP \textbf{02} (2004) 053}}, \href{https://arxiv.org/abs/hep-th/0401024}{[arXiv:hep-th/0401024]}.

\bibitem{2101.08875}
Donald Marolf, Jorge E. Santos, ``AdS Euclidean Wormholes'', {\hypersetup{urlcolor=vividviolet}\href{https://doi.org/10.1088/1361-6382/ac2cb7}{Class. Quant. Grav. \textbf{38} (2021) 22, 224002}}, \href{https://arxiv.org/abs/2101.08875}{[arXiv:2101.08875 [hep-th]]}.

\bibitem{0912.1061}
Sean A. Hartnoll, Joseph Polchinski, Eva Silverstein, David Tong, ``Towards Strange Metallic Holography'', {\hypersetup{urlcolor=vividviolet}\href{https://link.springer.com/article/10.1007\%2FJHEP04\%282010\%29120}{JHEP \textbf{04} (2010) 120}}, \href{https://arxiv.org/abs/0912.1061}{[arXiv:0912.1061 [hep-th]]}.

\bibitem{1910.06348}
Oscar Henriksson, Carlos Hoyos, Niko Jokela, ``Brane Nucleation Instabilities in Non-AdS/Non-CFT'', {\hypersetup{urlcolor=vividviolet}\href{https://link.springer.com/article/10.1007\%2FJHEP02\%282020\%29007}{JHEP \textbf{02} (2020) 007}}, \href{https://arxiv.org/abs/1910.06348}{[arXiv:1910.06348 [hep-th]]}.

\bibitem{2104.00022}
Raghu Mahajan, Donald Marolf, Jorge E. Santos, {\hypersetup{urlcolor=vividviolet}\href{https://link.springer.com/article/10.1007\%2FJHEP09\%282021\%29156}{JHEP \textbf{09} (2021) 156}}, ``The Double Cone Geometry Is Stable to Brane Nucleation'', \href{https://arxiv.org/abs/2104.00022}{[arXiv:2104.00022 [hep-th]]}.

\bibitem{0905.1180}
Brett McInnes, ``Bounding the Temperatures of Black Holes Dual to Strongly Coupled Field Theories on Flat Spacetime'', {\hypersetup{urlcolor=vividviolet}\href{https://iopscience.iop.org/article/10.1088/1126-6708/2009/09/048}{JHEP \textbf{09} (2009) 048}}, \href{https://arxiv.org/abs/0905.1180}{[arXiv:0905.1180 [hep-th]]}.

\bibitem{0910.4456}
Brett McInnes, ``Holography of the Quark Matter Triple Point'', {\hypersetup{urlcolor=vividviolet}\href{https://www.sciencedirect.com/science/article/abs/pii/S0550321310000994?via\%3Dihub}{Nucl. Phys. B \textbf{832} (2010) 323}}, \href{https://arxiv.org/abs/0910.4456}{[arXiv:0910.4456 [hep-th]]}.

\bibitem{1012.4056}
Brett McInnes, ``A Universal Lower Bound on the Specific Temperatures of AdS-Reissner-Nordstr\"om Black Holes with Flat Event Horizons'', {\hypersetup{urlcolor=vividviolet}\href{https://www.sciencedirect.com/science/article/abs/pii/S055032131100143X?via\%3Dihub}{Nucl. Phys. B \textbf{848} (2011) 474}}, \href{https://arxiv.org/abs/1012.4056}{[arXiv:1012.4056 [hep-th]]}.

\bibitem{1008.0231}
Brett McInnes, ``Fragile Black Holes'', {\hypersetup{urlcolor=vividviolet}\href{https://www.sciencedirect.com/science/article/abs/pii/S0550321310004372?via\%3Dihub}{Nucl. Phys. B \textbf{842} (2011) 86}}, \href{https://arxiv.org/abs/1008.0231}{[arXiv:1008.0231 [hep-th]]}.

\bibitem{0901.2574}
Keiju Murata, Tatsuma Nishioka, Norihiro Tanahashi, ``Warped AdS$_5$ Black Holes and Dual CFTs'', {\hypersetup{urlcolor=vividviolet}\href{https://academic.oup.com/ptp/article/121/5/941/2938707}{Prog. Theor. Phys. \textbf{121} (2009) 941}}, \href{https://arxiv.org/abs/0901.2574}{[arXiv:0901.2574 [hep-th]]}.

\bibitem{1409.3663}
Brett McInnes, ``A Holographic Bound on Cosmic Magnetic Fields'', {\hypersetup{urlcolor=vividviolet}\href{https://linkinghub.elsevier.com/retrieve/pii/S0550321315000036}{Nucl. Phys. B \textbf{892} (2015) 49}}, \href{https://arxiv.org/abs/1409.3663}{[arXiv:1409.3663 [hep-th]]}.

\bibitem{1201.6443}
Brett McInnes, ``Fragile Black Holes and an Angular Momentum Cutoff in Peripheral Heavy Ion Collisions'', {\hypersetup{urlcolor=vividviolet}\href{https://linkinghub.elsevier.com/retrieve/pii/S0550321312001782}{Nucl. Phys. B \textbf{861} (2012) 236}}, \href{https://arxiv.org/abs/1201.6443v2}{[arXiv:1201.6443 [hep-th]]}.

\bibitem{9706066}
S.O. Alexeyev, M.V. Pomazanov, ``Singular Regions in Black Hole Solutions in Higher Order Curvature Gravity'', \href{https://arxiv.org/abs/gr-qc/9706066}{[arXiv:gr-qc/9706066]}.


\bibitem{2106.08344}
Mario Herrero-Valea, ``The Shape of Scalar Gauss-Bonnet Gravity'', {\hypersetup{urlcolor=vividviolet}\href{https://link.springer.com/article/10.1007/JHEP03(2022)075}{JHEP \textbf{03} (2022) 075}}, \href{https://arxiv.org/abs/2106.08344}{[arXiv:2106.08344 [gr-qc]]}.

\bibitem{9903210}
Constantin P. Bachas, Pascal Bain, Michael B. Green, ``Curvature Terms in D-Brane Actions and Their M-Theory Origin'', {\hypersetup{urlcolor=vividviolet}\href{https://iopscience.iop.org/article/10.1088/1126-6708/1999/05/011}{JHEP \textbf{05} (1999) 011}}, \href{https://arxiv.org/abs/hep-th/9903210}{[arXiv:hep-th/9903210]}.


\bibitem{0904.4466}
Davide Gaiotto, Juan Maldacena, ``The Gravity Duals of $\mathcal{N}=2$ Superconformal Field Theories'', {\hypersetup{urlcolor=vividviolet}\href{https://link.springer.com/article/10.1007\%2FJHEP10\%282012\%29189}{JHEP \textbf{10} (2012) 18}}, \href{https://arxiv.org/abs/0904.4466}{[arXiv:0904.4466 [hep-th]]}.

\bibitem{9903210}
Constantin P. Bachas,  Pascal Bain, Michael B. Green, ``Curvature Terms in D-Brane Actions and Their M Theory Origin'', {\hypersetup{urlcolor=vividviolet}\href{https://iopscience.iop.org/article/10.1088/1126-6708/1999/05/011}{JHEP \textbf{05} (1999) 011}}, \href{https://arxiv.org/abs/hep-th/9903210}{[arXiv:hep-th/9903210]}.




\end{thebibliography}
\end{document}